\documentclass[useAMS,usenatbib]{mn2e}

\title[]{The giant star forming halo associated with the 
radio galaxy PKS1932-46.\thanks{Based on observations carried out at the
European Southern Observatory, Paranal, Chile with FORS2 on VLT (UT1).}}
\author[Villar-Mart\'\i n et al.]{M. Villar-Mart\'\i n$^{1}$\thanks{email:montse@iaa.es}, C. Tadhunter$^{2}$, R. Morganti$^{3}$, J. Holt$^{2}$\\
$^{1}$Instituto de Astrof\'\i sica de Andaluc\'\i a (CSIC), Aptdo. 3004, 18080 Granada, Spain\\
$^{2}$Dept. of Physics and Astronomy, University of Sheffield, Sheffield S3 7RH, UK\\
$^{3}$ASTRON, PO Box 2, 7990 Dwingeloo, the Netherlands} 
\begin{document}

\date{}

\pagerange{\pageref{firstpage}--\pageref{lastpage}} \pubyear{2002}

\maketitle

\label{firstpage}

\begin{abstract}
We report the discovery of a giant ($\sim$160 kpc) knotty
 extended emission line nebula
associated with the radio galaxy PKS1932-46 at $z=$0.23. The 2-d long slit
spectra, obtained with VLT-FORS2 at a large angle ($\sim$63$\degr$) to the 
radio source axis, shows
that the nebula extends
 well beyond the radio 
structure and the ionization cones of the active nucleus. This is one of the largest ionized nebulae yet 
detected around a radio galaxy at any redshift.  The analysis of the
ionization, morphological and kinematic properties of the knots suggests that these are star forming objects, probably  compact HII galaxies.
We propose that the giant structure  is a star forming halo associated with the
 debris of the merger that triggered the activity. This study reinforces the view that radio galaxies are activated by major mergers which also trigger substantial star formation. The star
formation activity  can extend on the scale of a galaxy group, beyond the old stellar halo of the host galaxy.

\end{abstract}


\section{Introduction}
Extended 
emission line haloes around radio galaxies have the potential to provide 
key information about the origins of the prodigious quasar and jet 
activity and how these objects are related to the evolution of 
giant elliptical galaxies (e.g. Baum \& Heckman 1989).
However, in 
order to use the emission lines in this way it is essential to distinguish 
the intrinsic properties of the warm ISM in the host galaxies/groups from 
the effects of the AGN/jet activity. Unfortunately, most previous 
long-slit spectroscopic studies have focussed on the  high 
surface brightness extended emission line regions (EELR) that are frequently aligned with the radio axis  (Baum \& Heckman 1989). Such structures are often dominated by 
jet-induced shocks (Clark et al. 1997, Villar-Mart\'\i n et al. 1999, Best, R\"ottgering \& Longair 2000) and it is difficult to determine the 
intrinsic properties of the ISM. 

Here we report observations of a 
spectacular emission line nebula in the FRII radio galaxy PKS1932-46 
($z=$0.23) that extends well beyond 
the radio structures and ionization cones associated with the nuclear activity, 
and allows us to investigate the properties of the undisturbed ISM in the 
extended halo around the host galaxy.

 Early spectroscopic observations of  PKS1932-46 
  revealed an emission line
nebulosity with a rich emission line spectrum extending out to a radius of 23$\arcsec$ ($\sim$92 kpc)
along the radio axis\footnote{We assume
$\Omega_{\Lambda}$=0.7, $\Omega_M$=0.3, H$_0$=65 km s$^{-1}$ Mpc$^{-1}$.
In this Cosmology, 1 arcsec = 4.0 kpc}  (Villar-Mart\'\i n et al. 1998, VM98 hereafter).  The object 
is surrounded by shell, filament and knot type structures
(VM98) that are typical of interacting systems.

All spectroscopic studies previously published of PKS1932-46 were based on shallower 
long slit
spectra taken along the radio axis. We present here deep optical VLT-FORS2 spectroscopy  with the slit at PA-9, misalinged by $\sim$63$\degr$ relative to the
radio axis (PA-72).
The observations and data reduction process are described in \S2.
The main results are described  in \S3 and the discussion and conclusions 
are 
presented in \S4.

\section[]{Observations}

Deep blue and red spectra were obtained using the 	
 FORS2 spectrograph on VLT (ESO-Paranal Observatory) on 24th Sept 2003. 
The log of the observations is presented in Table 1. The data were reduced following standard procedures
with STARLINK, IRAF and FIGARO packages.
The spectra were debiased, flat fielded, wavelength and flux calibrated, combined and cosmic ray removed, background subtracted, corrected for 
geometrical distortion and atmospheric
absorption features.  Comparison of several spectrophotometric standard stars taken with a 5$\arcsec$ slit throughout the run gave a relative flux calibration
accuracy of 5 per cent over the entire spectral range
of the observations.   The two resulting spectra (blue and red) were aligned spatially using the continuum centroids.

1D spectra were extracted from several apertures (ap. hereafter) shown in Fig.~1, which 
were selected to isolate individual knots (in most cases) along the slit. A pixel by pixel
analysis was performed in the $r\la$2$\arcsec$  region
(ap.10, Fig.~1). As an example, the spectra of ap. 4 and 6 are shown in Fig.~2.

The line profiles were fitted with 
Gaussian functions, which in general are
a good representation of the bulk of the line profiles.
The FWHM of the lines were corrected for the instrumental profile  (IP, Table 1) in quadrature.
The velocity shift $V_{sh}$ was calculated relative to the [OIII]$\lambda$5007 
central wavelength measured at the spatial position of the continuum
centroid. The spatial variation of the FWHM and $V_{sh}$  is shown in
Fig.3.  

\begin{figure}
\includegraphics{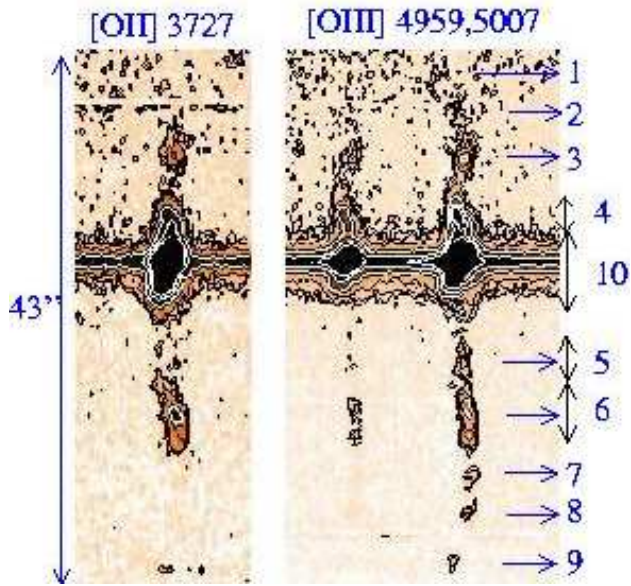}
\vspace{3.3in}
\caption{PKS1932-46: 2D spectra of the [OIII] and [OII] lines along PA -9$\degr.$ A spectacular extended region of ionized gas is detected 
 along $\sim$40$\arcsec$ ($\sim$160 kpc). It consists of a series of knots some of which are spatially unresolved.
 The apertures used in the analysis presented in this paper are indicated. }
\end{figure}

 The seeing for the observations was $\sim$0.92$\pm$0.03$\arcsec$, narrower than the 1.3$\arcsec$ 
wide slit. Point sources do not fill the slit. This could significantly affect especially the kinematic measurements of the compact knots (e.g. Villar-Mart\'\i n et al. 2003). Since the narrowness of the lines emitted by the knots has an important impact on our conclusions (see $\S$3), we have  been conservative
and assumed for the knots the minimum possible value of the IP, i.e., 
that of a point source ($\sim$6.1 \AA). Since some of the knots are spatially resolved, it is clear that the upper limits shown in Fig.~3 are very conservative.

\begin{table*}
\centering
\begin{tabular}{lllllllll}
\hline
 Date  & Instrument  &  Grating 	&  $\Delta\lambda$$^*$ & Exp. time & Slit  PA    &  IP &  Slit  width    & seeing \\ 
      &   \& Telescope &		&    \AA\  & sec     & (N to W) &  \AA\  & $\arcsec$ & $\arcsec$  \\ \hline
 24/09/2003  & UT1+FORS2          & GIRS\_600B   &	3400-6030	&    3x900     & -9$\degr$ &  6.5$\pm$0.2 & 1.3 &  $\sim$0.92$\pm$0.03  \\ 
 24/09/2003   & UT1+FORS2           & GRIS\_600RI  &	4950-8250 &    3x900     & -9$\degr$ &  7.4$\pm$0.2 & 1.3 &  $\sim$0.92$\pm$0.03 \\ 
\hline
\end{tabular}
\caption{Log of the observations. ($^*$Approximate useful spectral range) }
\end{table*}

\begin{figure*}
\includegraphics{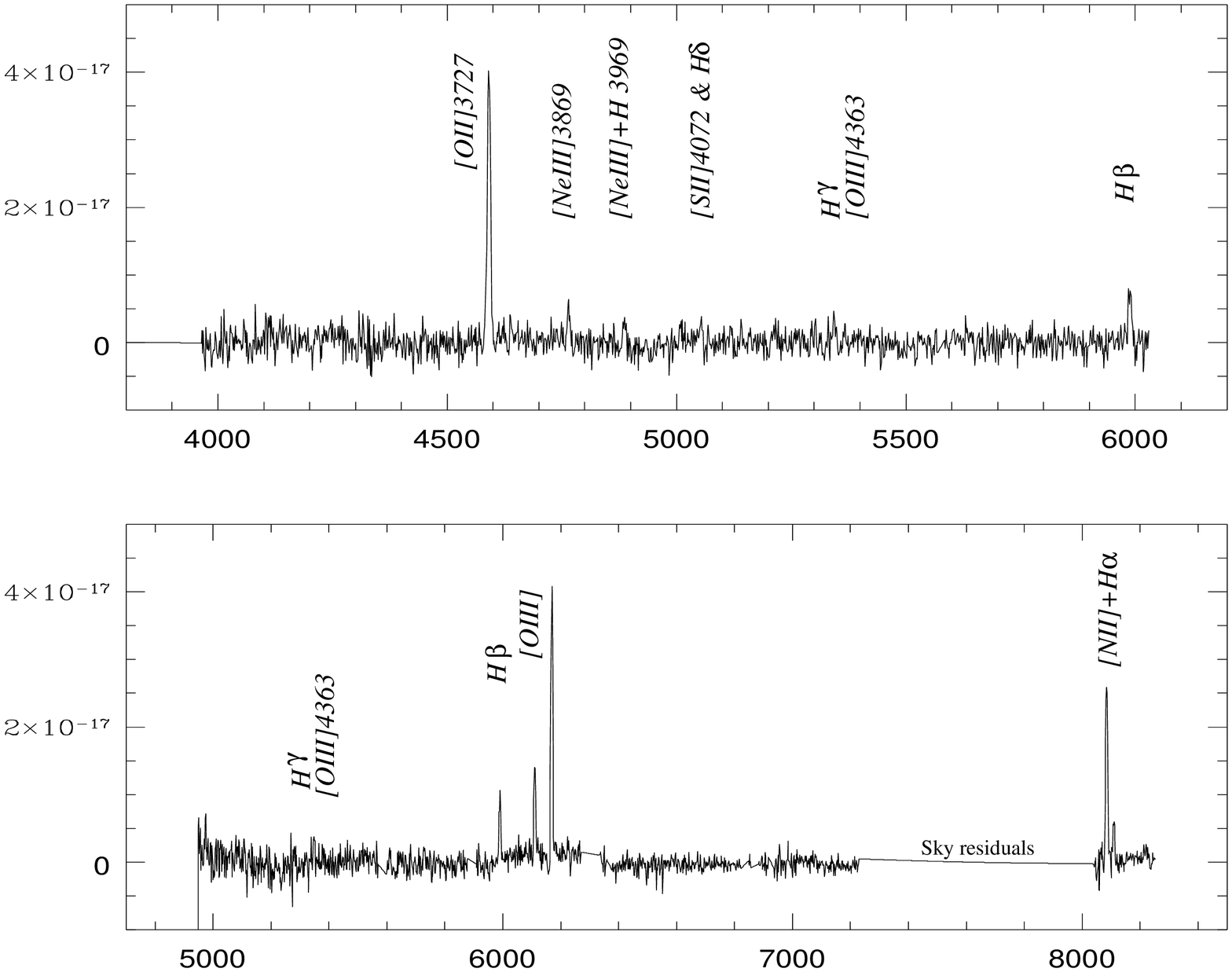}
\includegraphics{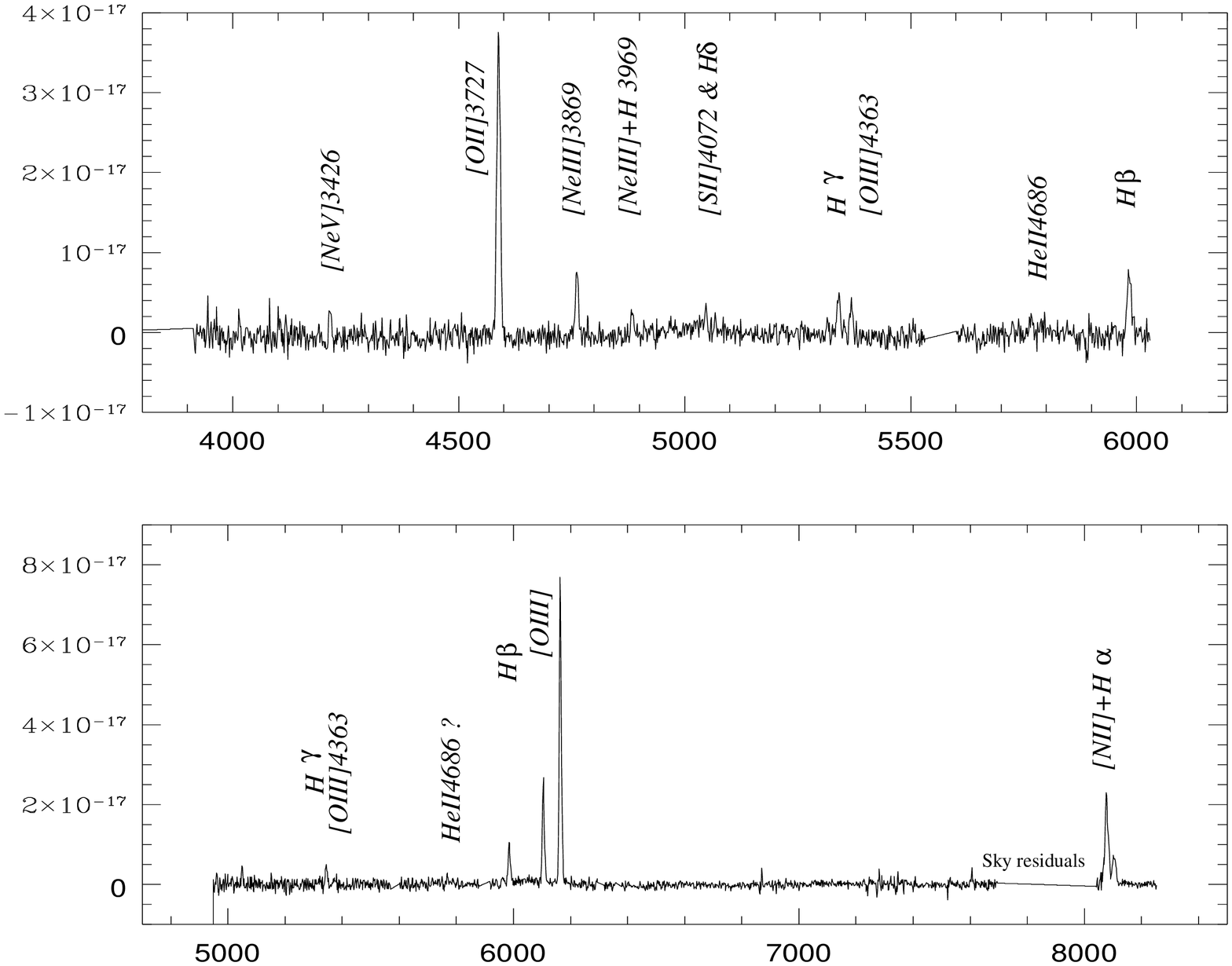}
\vspace{2.8in}
\caption{Blue and red spectra extracted from apertures 6 (left) and 4 (right).}
\end{figure*}

\section{Results}
Fig. 1 clearly reveals a giant knotty  emission line region that extends
for $\sim$40$\arcsec$ ($\sim$160 kpc),  well beyond the radio structures (VM98).
We have isolated three gaseous regions that differ in their kinematic 
 properties. 

\begin{itemize}

\item {\it The nuclear spectrum (r$<$1$\arcsec$, inner ap.10)}. This spectrum is typical of
the narrow line region (NLR) of active galaxies. It presents strong high
ionization lines ([NeV]$\lambda\lambda$3426,3446 and HeII$\lambda$4686)
as well as  relatively strong low ionization lines ([NII]$\lambda\lambda$6548,6583, [OI]$\lambda$6300,
[OII]$\lambda$3727, etc). The electron temperature measured using the [OIII] lines is very high with  $T_e$=20000$\pm$700 K.  

Fits to the [OIII] line profiles, which have the highest S/N, demonstrate 
that the best fit is obtained with two Gaussians
with FWHM 280$\pm$20 km  s$^{-1}$  and  1400$\pm$200
km  s$^{-1}$ respectively, with the broad component 
blueshifted by $\sim$150$\pm$100   km  s$^{-1}$ relative to the narrow component.

 The same kinematic components also 
provide good fits to all the other emission lines (similar to the 
situation in Cygnus A, for example: see Taylor, Tadhunter \& Robinson 2003), 
including the H$\alpha$+[NII] blend (see Figure 4). In the latter case, the 
[OIII] model provides a good fit to the wings of the blend without the 
need for the additional broad component (FWHM$\sim$2400 km  s$^{-1}$) 
which led to 
the broad line radio galaxy (BLRG) classification of VM98. 
Therefore, in contrast to the conclusions of these authors, the spectral properties 
of the nuclear regions are fully consistent with PKS1932-46 being a narrow 
line radio galaxy (NLRG). Further evidence against the idea that this is a 
broad line radio galaxy is provided by near-IR spectroscopic observations 
which show no signs of broad wings to the P$\alpha$ line (Bellamy, private 
communication). 

The measured H$\alpha$/H$\beta$ values are 3.8$\pm$0.2 and 6$\pm$2 for the narrow and the broad components respectively, implying significant reddening especially
for the broad component.

\begin{figure}
\includegraphics{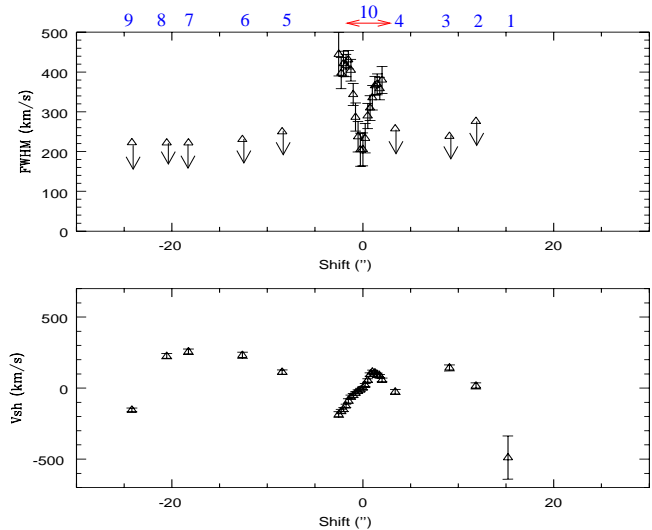}
\vspace{2.7in}
\caption{Spatial variation along the slit of the FWHM and velocity shift 
of [OIII]$\lambda$5007.
The numbers above the top  panel indicate the aperture nr. (see Fig.~1). FWHM values
for aperture 1 are not shown due to the noise. 
 The upper limits for the FWHM are very conservative values (see text). 
}
\end{figure}

\begin{figure}
\includegraphics{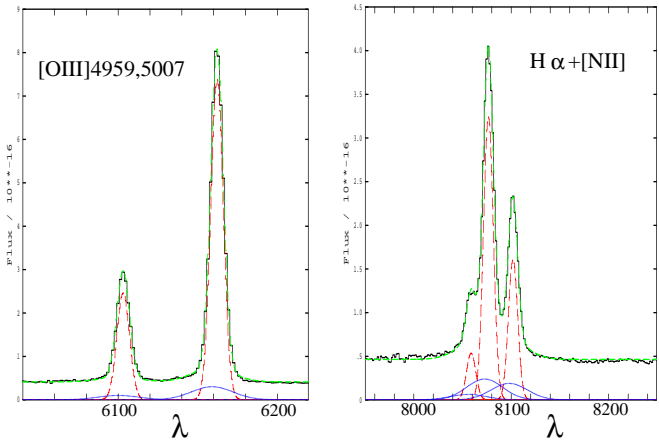}
\vspace{2.2in}
\caption{Fit to the [OIII] and H$\alpha$+[NII] lines.
The fit (dashed-green) and  data (solid-black) are shown in each panel above
the Gaussians of  the individual kinematics components, which have
been shifted down for clarity: narrow 
components (dashed-red) and broad components (solid-blue) {\it (see the electronic edition of the Journal
for a colour version of this figure)}. The same kinematic components found in the high S/N [OIII] lines 
provide good fits to the H$\alpha$+[NII] blend (see text).   This implies that PKS1932-46 is
a NLRG, rather than a BLRG as previously thought.}
\end{figure}

\item {\it The near nucleus (1$\arcsec$$<r<$3-5$\arcsec$, outer ap. 10 and ap. 4}). The spectrum of the extended gas in the near-nuclear region is characteristic of active galaxies,
with strong [NeV]$\lambda$3426 and HeII$\lambda$4686 and low ionization
lines (see Fig.~2). High electron temperatures are also measured 
with $T_e$ in the range 20000-24000 K in ap. 4 and the gas
just outside the nuclear region. The measured H$\alpha$/H$\beta$ = 3.1$\pm$0.2 is consistent with the  case B  value.
 
 The gas in this region  shows complex kinematics with broad
lines compared with the nuclear region (see Fig.~3) and  
two split narrow  components in aperture 4 (the emission in ap. 4 can be resolved
into two apparently spatially unresolved knots). 
These are shifted in velocity by 230$\pm$30 km  s$^{-1}$
and have FWHM([OIII]$\lambda$5007) = 220$\pm$40 and $\la$200 km  s$^{-1}$ 
respectively.  The integrated spectrum
of aperture 4 shows an  apparent anti-correlation between line width and ionization
level, similar to that found in other radio galaxies with clear signs of jet-gas
interactions (Clark et al. 1997, Villar-Mart\'\i n et al 1999). The [OIII] doublet lines have FWHM=280$\pm$20 km 
 s$^{-1}$ and FWHM of [NeV]$\lambda$3426)$\leq$250  km 
 s$^{-1}$. On the other hand, low ionization lines such as [NII], [OII] and the Balmer
lines have similar widths (FWHM=400-530 km s$^{-1}$) within the errors. 
 Complex kinematics is also obvious
in the 2D image of the long-slit spectrum in the H$\alpha$+[NII] region, with each line consisting of 
 two kinematic components. This kinematically perturbed region is about the
same size as the width of the radio cocoon in the direction perpendicular to the radio axis (VM98).  The kinematic and
ionization properties of this
gas are therefore likely to be strongly affected by 
shocks induced by the interaction between the gas and the radio structures.

\item {\it The extended halo (r$>$5$\arcsec$, ap. 5 to 9)}. 
This gas shows an irregular knotty morphology. The total extension is
striking ($\sim$40$\arcsec$ or $\sim$160 kpc). The individual knots have
sizes of $\la$10 kpc (the emission within aperture 6 can be resolved
into two different knots of spatial FWHM $\sim$1.4 $\arcsec$ and 2.0$\arcsec$ respectively). The lines (FWHM$\la$250  km s$^{-1}$) are
most probably narrower than in the 
inner regions (see \S2). The smoothness and symmetry of the velocity curve defined by the knots suggest that all of them follow a common kinematic pattern, which is decoupled from the gas at $r\la$5$\arcsec$ (Fig.~3).

The Balmer decrement was measured in ap. 3 and 6, giving
H$\alpha$/H$\beta$=2.7$\pm$0.5 and 3.2$\pm$0.3 respectively, consistent with
the case B value within the errors.

We show in Fig. 5 three diagnostic diagrams involving the main emission
lines that could be detected in most apertures across the extended
halo. We have also included  ap. 4 (discussed above),
 the integrated nuclear spectrum and the 
  gas extended {\it along the radio axis} discussed by VM98  (inner and outer EELR).
 For comparison, the standard solar metallicity  power-law (index $\alpha$=-1.5), single slab photoionization
model
sequence often applied to low redshift active galaxies (e.g. Robinson et al. 1987) is also shown. The ionization parameter $U$ varies along the sequence 
(models with log$U$=-3, -2, -1 are marked). The position of a sample of  HII galaxies extracted
from the catalogue by Terlevich et al. (1991) is also shown. 
The diagrams show that  the ionization properties of the knots 
vary along the slit. The main results are:
\begin{itemize}

\item Ap. 3, 4 and 5 and the nuclear spectrum (open symbols) lie close to the AGN photoionization models.
The different positions  of the knots can be naturally explained by a decrease in the ionization level with increasing distance from the AGN:  ap. 3 and 5
(which are located at similar distance (8-10$\arcsec$ from the AGN) show similar ionization level, while the gas in ap. 4 ($\sim$3.5$\arcsec$ from the AGN) has higher
ionization level. As discussed above, it is also possible that the gas in ap. 4 (see above) is
partially ionized by shocks.  

 The outer EELR  (VM98) is gas located at $\sim$20$\arcsec$ {\it along the radio axis} and just beyond the Eastern radio lobe.
It follows the same $U$ sequence discussed above and presents the highest ionization level. Since it is apparently  located at a
much larger distance than ap. 3 to 5 ($r<$10$\arcsec$), 
this  suggests a change in the physical conditions (lower
density) or that the AGN continuum is stronger along the radio axis (see \S4). 

\item Ap. 6 shows a mixed spectrum. The [NII] emission places it in the area of the star forming objects while the oxygen lines
 place it closer to the AGN models.

\item  The outer apertures (2, 7 and 8+9) show a spectrum characteristic of star
forming objects. The three apertures have very similar [OIII]/H$\beta\sim$3-3.5. 

\end{itemize}

It seems therefore, that there is a gradual change on the balance of the
ionization mechanism when we move outwards along the slit. AGN related
processes dominate
 in the inner regions   ($r\la$8$\arcsec$ or
ap. 3, 4, 10 and 5). A mixture of stellar and AGN 
photoionization is then apparent at intermediate distances 
($\sim$13$\arcsec$, ap. 6) and at larger distances  
the objects are ionized by stars (ap. 2, 7, 8, 9).

 The H$\beta$ luminosities of the outer knots (2, 7, 8, 9) are
$\sim$1-2$\times$10$^{39}$ erg s$^{-1}$, a factor between
$\sim$2 and 10 times lower than the more  interior knots, and within the range
of values measured in  HII galaxies (Terlevich et al. 1991). The implied star forming
rates are $\sim$0.02-0.05 M$_{\odot}$ yr$^{-1}$, ignoring underlying line absorption
and reddening (Kennicutt 1998).  Their compact appearance, small sizes ($\la$10 kpc) and
narrow emission lines
(FWHM$\la$250 km s$^{-1}$) 
 are all consistent with values measured in  compact HII galaxies (e.g. Terlevich et al. 1991,
Guzm\'an et al 1997).
Continuum is detected from ap. 8 (and maybe 9 as well), although its
nature is not known
(nebular or stellar). The non detection of continuum in  ap.
2 and 7 does not contradict the stellar
interpretation, since large H$\beta$ equivalent widths ($>$100 \AA\ rest frame) are expected for young stellar ages ($\la$few Myr) 
 (e.g Stasinska, Schaerer \& Leitherer 2001) 
which would imply continuum levels well under the detection limit.
 The [OII], [OIII] and H$\beta$ lines were used to estimate the oxygen abundance  (Aller 1984) of 
apertures 2,  7 and 8+9.  A  $T_e$ value of 12000 K was assumed (higher temperatures  produce lower abundances).
The derived abundances are $\sim$18\%, $\la$21\% and 15\%  (O/H)$_{\odot}$ 
for apertures 2, 7 and 8+9 respectively. Similar values are  often derived for compact HII galaxies  (e.g. Terlevich et al. 1991).

All properties of the outer knots studied here are therefore consistent with compact HII galaxies.
Although the spectra of the inner knots (ap. 3 to 6) are distorted by AGN related processes,  their compact appearance, small sizes ($\la$10 kpc) and
narrow emission lines
(FWHM$\la$250 km s$^{-1}$) 
 are also consistent with values measured in compact HII galaxies. 
Therefore, we propose that {\it all the knots (ap. 2 to 9)} in the giant halo are star forming objects

\begin{figure*}
\includegraphics{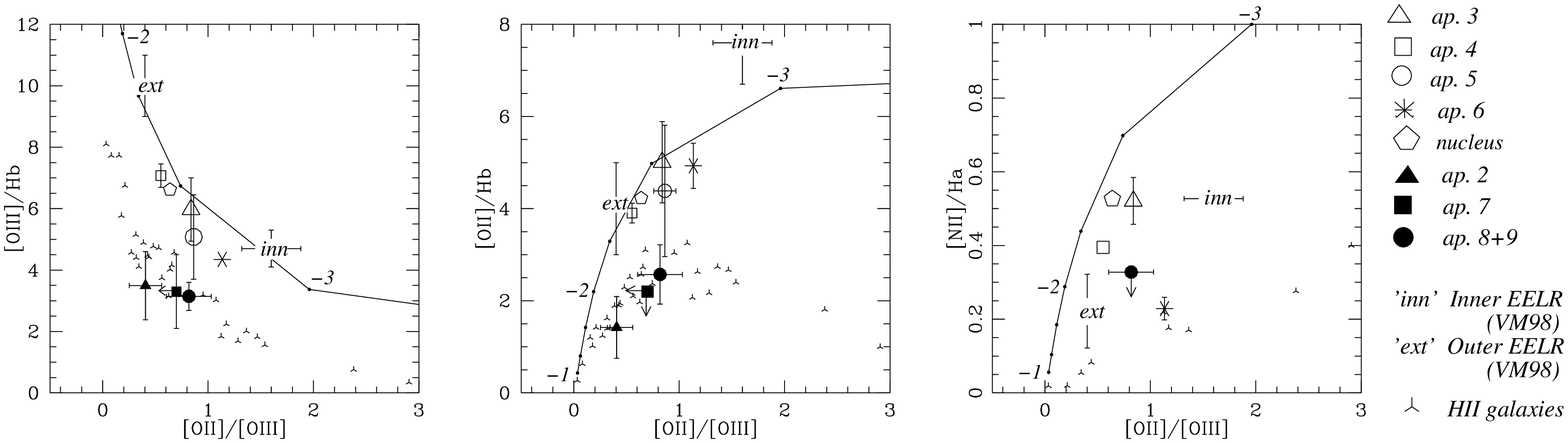}
\vspace{2.2in}
\caption{Diagnostic diagrams showing the position of all apertures where the main lines could be measured. The standard power law ($\alpha$=-1.5)  ionization parameter
$U$ sequence often
applied to radio galaxies
is  shown as a solid black line (Robinson et al. 1987). The negative numbers 
mark log($U$).  A sample of HII galaxies are also plotted (Terlevich et al. 1991), as well as the line ratios measured in the extended gas along the radio
axis by VM98 ({\it inn} and {\it ext}).
Errorbars smaller than the symbols size
are not shown.  [NII]/H$\alpha$ could not be measured for ap. 2 and 5. The gas within ap. 3, 4 and 5 (open symbols) is likely to be mostly
ionized by  AGN related processes. Moving outward along the slit, ap.
6 (star symbol) shows mixed properties between AGNs (left and middle panels) and star
forming galaxies (right panel). The outer knots 
 (ap. 2, 7 and 8+9, solid symbols) have spectroscopic
properties consistent with  HII galaxies. There is therefore
a gradual change on the ionization mechanism along the slit such that stellar
ionization becomes dominant in the outer knots.}
\end{figure*}
\end{itemize}

\section{Discussion and conclusions}

We have discovered a giant gaseous knotty structure
that extends for a striking distance of $\sim$160 kpc. This is one of the largest
emission line halos ever detected around an active galaxy at any redshift.
We have proposed that the halo comprises of star forming objects, probably compact HII galaxies.

Based on continuum properties, young stellar populations (YSP) have been detected in the nuclear regions of a significant fraction of nearby radio galaxies
 (e.g. Tadhunter et al. 2002,  2005). However, most of these YSP represent relatively old post-starburst populations (0.05-2 Gyr).
 Our new observations show that the star formation is ongoing in PKS1932-46 and it can actually spread over spatial scales of more than 100 kpc. 
Since the object lies in a rich  environment where a merger/interaction event is likely to be taking place (VM98), we propose that the giant halo is a residual feature of such process where compact star forming objects  have formed. 
Morphologically, the giant halo
reminds us of the linear knotty ionized structure extended for
$\ga$tens of kpc  associated with the system IC 1182, which shows  clear signs of 
interactions and/or mergers (e.g. Moles et al. 2004). 
 The authors 
propose that some of the H$\alpha$ condensations 
found  in the large scale tails  could
be tidal dwarf proto-galaxies. The existence of isolated intergalactic 
HII regions  has also been reported at $\sim$100 kpc  from the early type galaxy NGC1490 (Oosterloo et al. 2004). The HII regions are associated with large HI clouds lying along a 500 kpc long arc, which is likely to be a product of tidal interactions or merger events.

Giant kinematically unperturbed halos ($\ga$100 kpc) of ionized gas have been found in some high redshift radio galaxies ($z\sim$2.5, e.g.  Villar-Mart\'\i n et al. 2003). Unfortunately, the slit was always located along the radio axis and  the observed properties are strongly distorted by the nuclear activity. 
It is interesting to consider  whether the 
 giant halo discovered around PKS1932-46 has a similar nature. The detailed study presented here opens the possibility that the high redshift
halos   are also giant star forming regions, the product of ongoing merger events. Star formation across giant spatial scales (several tens of kpc to 1 Mpc)  has been reported for several radio galaxies at $z\sim$2-4 (Stevens et al. 2003)
based on sub-mm observations.

We have concluded that AGN photoionization  is likely to excite the gas in
the knots located closer to the  nucleus (ap. 3, 4, 5). Jet-induced shocks could
also play a role in ap.4 (see \S3).
The slit was placed at -9$\degr$, 
shifted
by $\sim$63$\degr$  relative to the radio axis. 
In the simplest scenario, the radio axis is
 also the axis of the ionization cones, which are expected to have 
an opening angle of 90$\degr$ (Barthel 1989). In such case,
and given that PKS1932-46 is a NLRG (see \S3)
  the slit position 
is outside the ionization cones.  A misalignment of $\ga$15$\degr$ between the
radio axis and the cones axis is required
for AGN photoionization to be plausible. Otherwise we need to invoke
larger cone opening angles, or the existence of a porous torus that allows
some filtered continuum to escape in directions outside the standard 90$\degr$
cones.

Overall, this study reinforces the view that the nuclear activity in
radio galaxies is triggered by major galaxy mergers and there is  substantial
associated star formation  (e.g. Heckman et al. 1986), which can extend on the scale of a galaxy group, beyond the old stellar halo of the host galaxy.

\section*{Acknowledgments}
We thank the staff on Paranal for their support. MVM is supported
by the Spanish National program Ram\'on y Cajal. JH
acknowledges support from a PPARC studentship. We thank the referee
Andy Robinson for useful
comments.

\end{document}